\documentclass{article}%
\usepackage{amsfonts}
\usepackage{amsmath}
\usepackage{hyperref}
\usepackage{amssymb}
\usepackage{graphicx}%
\begin{document}

\title{Mean Sinc Sums and Scale Invariant Scattering}
\author{Thomas Curtright\thanks{{\footnotesize curtright@miami.edu}}\medskip\\Department of Physics, University of Miami, Coral Gables, FL 33124}
\maketitle

\begin{abstract}
\textit{Scattering from a scale invariant potential in two spatial dimensions
leads to a class of novel identities involving the sinc} \textit{function.}

\end{abstract}

It is well-known that non-relativistic scattering by a $V=\kappa/r^{2}$
potential is scale invariant \cite{Jackiw}, in any number of spatial
dimensions, and the scattering problem is well-defined mathematically
\cite{Newton} for all $\kappa>0$. \ However, when computed classically, the
total (integrated) scattering cross section $\sigma$ is \emph{infinite}
(\cite{Newton}, \S 5.6). \ An infinite $\sigma$ is also obtained for quantum
mechanical scattering by an inverse square potential in three spatial
dimensions \cite{3DRemark} (a property shared by Coulomb scattering).

On the other hand, in two spatial dimensions the integrated cross section,
$\sigma=\int_{0}^{2\pi}\left(  \frac{d\sigma}{d\theta}\right)  d\theta$, is
\emph{finite} when computed using quantum mechanics. \ The result for a
mono-energetic beam is
\begin{equation}
\sigma=\frac{2\pi^{2}m\kappa}{\hbar^{2}k} \label{SimpleSigma}%
\end{equation}
where the incident energy is $E=\hbar^{2}k^{2}/\left(  2m\right)  $. \ This
result follows from a straightforward application of phase-shift analysis for
the potential $V=\kappa/r^{2}$, upon realizing a peculiar identity involving
the sinc function, $\operatorname{sinc}\left(  z\right)  \equiv\sin\left(
z\right)  /z$. \ A succinct form of the identity in question is
\begin{equation}
1=\frac{\sin\left(  \pi x\right)  }{\pi x}+2\sum_{l=1}^{\infty}\frac{\left(
-1\right)  ^{l}\sin\left(  \pi\sqrt{l^{2}+x^{2}}\right)  }{\pi\sqrt
{l^{2}+x^{2}}} \label{ID1}%
\end{equation}
Note the \textquotedblleft mean\textquotedblright\ (rms) arguments.
\ Remarkably, all higher powers of $x$ cancel when terms on the RHS are
expanded as series in $x^{2}$, as a consequence of familiar $\zeta\left(
2n\right)  $ exact values for integer $n>0$.

In more detail, the scattering amplitude for a plane wave incident on a
potential, in two spatial dimensions, is
\begin{equation}
f\left(  \theta\right)  =\sqrt{\frac{2}{\pi k}}\sum_{l=-\infty}^{\infty
}e^{il\theta}e^{i\delta_{l}}\sin\left(  \delta_{l}\right)
\end{equation}
For the case at hand the phase shifts are given exactly by%
\begin{equation}
\delta_{l}=\frac{\pi}{2}\left(  \sqrt{l^{2}}-\sqrt{l^{2}+2m\kappa/\hbar^{2}%
}\right)  \label{PhaseShifts}%
\end{equation}
with no $k$ dependence, thereby exhibiting scale invariance in this context
\cite{CutoffRemark}. \ The differential cross section is of course
$d\sigma/d\theta=\left\vert f\left(  \theta\right)  \right\vert ^{2}$ which
integrates to give the total cross section%
\begin{equation}
\sigma=\frac{4}{k}\sum_{l=-\infty}^{\infty}\sin^{2}\left(  \delta_{l}\right)
=\frac{4}{k}\left(  \sin^{2}\left[  \frac{\pi}{2}\sqrt{2m\kappa/\hbar^{2}%
}\right]  +2\sum_{l=1}^{\infty}\sin^{2}\left[  \frac{\pi}{2}\left(
\sqrt{l^{2}+2m\kappa/\hbar^{2}}-l\right)  \right]  \right)  \label{Unsummed}%
\end{equation}
The final result for $\sigma$ therefore follows from the evaluation of this
last sum. \ 

\ \bigskip

It is indeed pleasing to find%
\begin{equation}
\frac{\pi^{2}x^{2}}{4}=\sin^{2}\left(  \frac{\pi x}{2}\right)  +2\sum
_{l=1}^{\infty}\sin^{2}\left(  \frac{\pi}{2}\left(  \sqrt{l^{2}+x^{2}%
}-l\right)  \right)  \label{ID2}%
\end{equation}
Convergence of the sum follows by Raabe's test. \ But it is very surprising
that the \emph{net} contribution of the RHS is only the leading term in the
expansion $\sin^{2}\left(  \frac{\pi x}{2}\right)  =\frac{\pi^{2}x^{2}}%
{4}+O\left(  x^{4}\right)  $. \ 

As a check on (\ref{ID2}), expand each term under the sum, $\sin^{2}\left(
\frac{\pi}{2}\left(  \sqrt{l^{2}+x^{2}}-l\right)  \right)  =\frac{1}{16}%
\frac{\pi^{2}x^{4}}{l^{2}}+O\left(  x^{6}\right)  $. \ So to leading order the
sum gives $2\sum_{l=1}^{\infty}\frac{1}{16}\frac{\pi^{2}x^{4}}{l^{2}}=\frac
{1}{48}\pi^{4}x^{4}$ upon using $\zeta\left(  2\right)  =\frac{1}{6}\pi^{2}$.
\ This exactly cancels the next to leading order from the first term on the
RHS of (\ref{ID2}), namely, $\sin^{2}\left(  \frac{\pi x}{2}\right)  =\frac
{1}{4}\pi^{2}x^{2}-\frac{1}{48}\pi^{4}x^{4}+O\left(  x^{6}\right)  $. \ And so
it goes to higher orders in powers of $x^{2}$, as may be verified by computer,
say to $O\left(  x^{100}\right)  $. \ A formal, perhaps convincing argument
that the result is true to all orders in $x^{2}$ is obtained by interchanging
the sum over $l$ with the series expansion sums for the various $\sin
^{2}\left(  \frac{\pi}{2}\left(  \sqrt{l^{2}+x^{2}}-l\right)  \right)  $,
regulating the resulting divergent sums over $l$ by analytic continuation of
the $\zeta$ function \cite{ZetaReg}, and using $\zeta\left(  -2n\right)  =0$
for all integer $n>0$.

Given that (\ref{ID2}) holds for all real $x$, differentiation or integration
produces a set of related results. \ In particular, $\frac{2}{\pi^{2}x}%
\frac{d}{dx}$ applied to both LHS and RHS of (\ref{ID2}) gives (\ref{ID1}).
\ Alternatively, (\ref{ID1})\ may be obtained directly using the steps shown
in Appendix A. \ Integration of $x\times$(\ref{ID1}) then gives (\ref{ID2}).
\ Then again, differentiating (\ref{ID1}) leads to an identity that can be
easily established by a Sommerfeld-Watson transformation, as shown in Appendix
B. \ The results exhibited in (\ref{ID1}) and (\ref{ID2}), and those in the
related set, are not extant in the literature, so far as I have been able to
determine, although an engaging survey of other surprising results involving
the sinc function can be found in \cite{Sinc}.

In closing, a few brief remarks are warranted about the simple form for
$\sigma$ as given by (\ref{SimpleSigma}). \ Since $m\kappa/\hbar^{2}$ is a
dimensionless parameter, a priori it would be allowed on dimensional grounds
to have $\sigma=f\left(  m\kappa/\hbar^{2}\right)  /k$ where the function $f$
need not be linear. \ Indeed, the sum in (\ref{Unsummed}) is exactly of this
form before any simplification. \ The underlying physical reason that sum
actually turns out to be linear in $\kappa$ is not very clear, and requires
further examination. \ This matter is under study.\bigskip

\textbf{Acknowledgements} \ I thank C. Bender and A. Turbiner for discussions
and encouragement to write up these results, and I thank T.S. Van Kortryk for
checking some of the math. \ I received financial support from the United
States Social Security Administration.

\newpage

\subsection*{Appendix A}

Here are some steps leading to (\ref{ID1}). \ First, expand the summand as a
power series in $x^{2}$.
\begin{equation}
\frac{\sin\left(  \pi\sqrt{l^{2}+x^{2}}\right)  }{\pi\sqrt{l^{2}+x^{2}}}%
=\sum_{n=0}^{\infty}\frac{\left(  x^{2}\right)  ^{n}}{n!2^{n}}\left(  \frac
{1}{l}\frac{d}{dl}\right)  ^{n}\frac{\sin\left(  \pi l\right)  }{\pi l}
\tag{A1}\label{A1}%
\end{equation}
Next, use the well-known relation expressing spherical Bessel functions in
terms of the sinc function,
\begin{equation}
j_{n}\left(  z\right)  =\left(  -z\right)  ^{n}\left(  \frac{1}{z}\frac{d}%
{dz}\right)  ^{n}\frac{\sin z}{z} \tag{A2}\label{A2}%
\end{equation}
to obtain%
\begin{equation}
\frac{\sin\left(  \pi\sqrt{l^{2}+x^{2}}\right)  }{\pi\sqrt{l^{2}+x^{2}}}%
=\sum_{n=0}^{\infty}\frac{\left(  x^{2}\right)  ^{n}\pi^{n}}{n!2^{n}}%
\frac{j_{n}\left(  \pi l\right)  }{\left(  -l\right)  ^{n}}=\sum_{n=1}%
^{\infty}\frac{\left(  -\pi x^{2}\right)  ^{n}}{n!2^{n}}\frac{1}{l^{n}}%
\sqrt{\frac{1}{2l}}J_{n+1/2}\left(  \pi l\right)  \tag{A3}\label{A3}%
\end{equation}
Note that the $n=0$ term vanishes since $j_{0}\left(  \pi l\right)
\propto\sin\left(  \pi l\right)  =0$ for integer $l$. \ Performing the sum
over $l$ before the sum over $n$ then leads to
\begin{equation}
\sum_{l=1}^{\infty}\frac{\left(  -1\right)  ^{l}}{l^{n+1/2}}~J_{n+1/2}\left(
\pi l\right)  =\frac{-\pi^{n}/\sqrt{2}}{\left(  2n+1\right)  !!}\text{ \ \ for
integer \ \ }n\geq1 \tag{A4}\label{A4}%
\end{equation}
where $k!!$ is the double factorial. \ Verification of (\ref{A4}) is left as
an exercise for the reader. \ It boils down to the following identity
involving the Bernoulli numbers.%
\begin{equation}
\frac{1}{\left(  2n+1\right)  !!}=\left(  -2\right)  ^{n+1}\sum_{k=0}^{n}%
\frac{B_{n+k+1}}{k!\left(  n-k\right)  !\left(  n+k+1\right)  }\text{ \ \ for
integer }n\geq1\text{.} \tag{A5}\label{A5}%
\end{equation}
(NB \ This identity also holds for $n=0$ provided that $B_{1}=-1/2$.)
\ Finally, the sum over $n$ gives%
\begin{equation}
2\sum_{l=1}^{\infty}\frac{\left(  -1\right)  ^{l}\sin\left(  \pi\sqrt
{l^{2}+x^{2}}\right)  }{\pi\sqrt{l^{2}+x^{2}}}=-2\sum_{n=1}^{\infty}%
\frac{\left(  -\pi x^{2}\right)  ^{n}}{n!2^{n+1/2}}\frac{\pi^{n}/\sqrt{2}%
}{\left(  2n+1\right)  !!}=-\sum_{n=1}^{\infty}\frac{\left(  -\pi^{2}%
x^{2}\right)  ^{n}}{\left(  2n+1\right)  !}=1-\frac{\sin\left(  \pi x\right)
}{\pi x} \tag{A6}\label{A6}%
\end{equation}
and hence the result (\ref{ID1}).

\newpage

\subsection*{Appendix B}

Differentiating with respect to $x$ the summand on the RHS of (\ref{ID1})
gives a combination of cosine and sinc functions.
\begin{equation}
\frac{d}{dx}\frac{\left(  -1\right)  ^{l}\sin\left(  \pi\sqrt{l^{2}+x^{2}%
}\right)  }{\pi\sqrt{l^{2}+x^{2}}}=x\left(  \frac{\left(  -1\right)  ^{l}%
}{l^{2}+x^{2}}\right)  \left(  \cos\left(  \pi\sqrt{l^{2}+x^{2}}\right)
-\frac{\sin\left(  \pi\sqrt{l^{2}+x^{2}}\right)  }{\pi\sqrt{l^{2}+x^{2}}%
}\right)  \tag{B1}\label{B1}%
\end{equation}
Summing either of these two terms can be done by a Sommerfeld-Watson
transformation. \ Upon replacing $l$ with the complex variable $z$, neither
term has a cut in $z$ since both cosine and sinc are even functions of their
arguments. \ After division by $\sin\left(  \pi z\right)  $, either term falls
off as $1/\left\vert z\right\vert ^{2}$ for large $\left\vert z\right\vert $.
\ By the residue theorem then,%
\begin{equation}
0=%
{\displaystyle\oint_{C}}
\frac{\cos\left(  \pi\sqrt{z^{2}+x^{2}}\right)  }{z^{2}+x^{2}}\frac{1}{\sin\pi
z}dz=\frac{1}{2\pi i}\left(  \sum_{l=-\infty}^{\infty}\left(  -1\right)
^{l}\frac{\cos\left(  \pi\sqrt{l^{2}+x^{2}}\right)  }{l^{2}+x^{2}}-\frac{\pi
}{x\sinh\pi x}\right)  \tag{B2}\label{B2}%
\end{equation}
where the contour $C$ is the infinite radius limit of a circle centered on the
origin. \ The sum $\sum_{l=-\infty}^{\infty}$ on the RHS of (\ref{B2}) is the
contribution of the residues at the poles in $1/\sin\left(  \pi z\right)  $
for integer $z$, while the $\frac{-\pi}{x\sinh\pi x}$ term is the contribution
from the poles in $1/\left(  z^{2}+x^{2}\right)  $ at $z=\pm ix$ for real $x$.
\ Thus
\begin{equation}
\frac{\pi}{\sinh\pi x}=x\sum_{l=-\infty}^{\infty}\left(  -1\right)  ^{l}%
\frac{\cos\left(  \pi\sqrt{l^{2}+x^{2}}\right)  }{l^{2}+x^{2}} \tag{B3}%
\label{B3}%
\end{equation}
The same contour integral analysis gives the identical result
\begin{equation}
\frac{\pi}{\sinh\pi x}=x\sum_{l=-\infty}^{\infty}\frac{\left(  -1\right)
^{l}}{l^{2}+x^{2}}\frac{\sin\left(  \pi\sqrt{l^{2}+x^{2}}\right)  }{\pi
\sqrt{l^{2}+x^{2}}} \tag{B4}\label{B4}%
\end{equation}
Therefore the RHS of (\ref{ID1}) is independent of $x$.
\begin{equation}
0=\frac{d}{dx}\left(  \sum_{l=-\infty}^{\infty}\frac{\left(  -1\right)
^{l}\sin\left(  \pi\sqrt{l^{2}+x^{2}}\right)  }{\pi\sqrt{l^{2}+x^{2}}}\right)
\tag{B5}\label{B5}%
\end{equation}
Evaluation of this last sum at $x=0$ then verifies (\ref{ID1}).


\begin{thebibliography}{9}
\bibitem {3DRemark}This follows from partial wave analysis similar to the 2D
case, as given in the text, except in 3D the phase shift is $\delta_{l}%
=\frac{\pi}{2}\left(  \sqrt{l\left(  l+1\right)  }-\sqrt{l\left(  l+1\right)
+2m\kappa/\hbar^{2}}\right)  $ and the sum $\sum_{l=0}^{\infty}\left(
2l+1\right)  \sin^{2}\delta_{l}$ is then divergent because of the additional
$2l+1$ multiplicity factor.
\bibitem {CutoffRemark}If the potential is cut-off outside some large finite
radius, $R$, such that $V=\kappa\Theta\left(  R-r\right)  /r^{2}$ where
$\Theta$ is the Heaviside step function, then the scattering amplitudes are
given in terms of Bessel functions as a ratio of Wronskians, $e^{2i\delta_{l}%
}=-W\left[  J_{\nu\left(  l\right)  }\left(  kR\right)  ,H_{l}^{\left(
2\right)  }\left(  kR\right)  \right]  /W\left[  J_{\nu\left(  l\right)
}\left(  kR\right)  ,H_{l}^{\left(  1\right)  }\left(  kR\right)  \right]  $,
where $\nu\left(  l\right)  =\sqrt{l^{2}+2m\kappa/\hbar^{2}}$, with the result
(\ref{PhaseShifts}) recovered in the limit $kR\rightarrow\infty$.
\end{thebibliography}
\end{document}